# Physical Mechanisms of Interaction of Cold Plasma with Polymer Surfaces

Edward Bormashenko [a,b*] Gene Whyman [a], Victor Multanen [b], Evgeny Shulzinger [a], Gilad Chaniel [a,c]

[a]Ariel University, Physics Faculty, 40700, P.O.B. 3, Ariel, Israel
[b]Ariel University, Chemical Engineering and Biotechnology Faculty, 40700, P.O.B. 3, Ariel, Israel
[c]Bar Ilan University, Physics Faculty, 52900, Ramat Gan, Israel

E-mail: edward@ariel.ac.il

**Abstract**

Physical mechanisms of the interaction of cold plasmas with organic surfaces are discussed. Trapping of plasma ions by the $CH_2$ groups of polymer surfaces resulting in their electrical charging is treated. Polyethylene surfaces were exposed to the cold radiofrequency air plasma for different intervals of time. The change in the wettability of these surfaces was registered. The experimentally established characteristic time scales of the interaction of cold plasma with polymer surfaces are inversely proportional to the concentration of ions. The phenomenological kinetic model of the electrical charging of polymer surfaces by plasmas is introduced and analyzed.

1.  **Introduction**

Plasma treatment (low and atmospheric-pressure) is broadly used for modification of the surface properties of polymer materials [1-6]. The plasma treatment creates a complex mixture of surface functionalities which influence physical and chemical properties of polymers, and results in a dramatic change in the wetting behaviour of the surface; this is true for both solid and liquid surfaces [7-17].

Not only the chemical structure but also the roughness of the surface is affected by the plasma treatment, which also could change the wettability of the surface [18]. Plasma treatment usually strengthens the hydrophilicity of treated polymer surfaces. However, the surface hydrophilicity created by plasma treatment is often lost over time. This effect of decreasing hydrophilicity is called "hydrophobic recovery" [19-28]. The phenomenon of hydrophobic recovery is usually attributed to a variety of physical and chemical processes; however, their influence remains highly debatable [19-28].

Our investigation is focused on the interaction of so-called "cold" plasma discharges with organic-polymer surfaces which are generally solid or liquid. The "cold" plasma discharges are characterized by the temperatures of electrons $T_e \cong 1-10\,\text{V}$ and temperatures of ions which are much lower, than those of electrons $T_i \ll T_e$ (recall that the temperature associated with $T=1\text{V}$ equals 11605K). The concentration of charge carriers in cold plasmas is typically $n \approx 10^{14}-10^{19}\,\text{m}^{-3}$ (see Ref. 29).

Plasmas are joined to solid surfaces across thin positively charged layers called sheaths [29], depicted in Fig.1. The origin of plasma sheaths may be understood on account that the electron thermal velocity is much larger than the ion thermal velocity. Thus, at the surface bounding cold plasma, a potential exists to contain the more mobile charged species [29]. This allows the flow of positive and negative carriers to the wall to be balanced. In the usual situation of the plasma, consisting of equal numbers of positive ions and electrons, the electrons are far more mobile than the ions. The plasma will therefore be charged positively with the respect to a grounded wall. Plasma sheaths accelerate ions and it is reasonable to attribute the modification of organic surfaces by plasmas to the collisions of ions accelerated by the electric field of a sheath with the moieties constituting an organic surface. The electrons of the sheath practically do not transfer energy to much heavier organic moieties forming the surface [29]. The typical thickness of plasma sheath is on the order of magnitude of the Debye length [29]. In spite of the great theoretical and experimental effort spent on the investigation of the interaction of cold plasmas with organic surfaces, the mechanism of this interaction remains obscure. Our study introduces these mechanisms as verified by the experimental research, thus validating the theoretical assumptions.

## 2. Experimental

Smooth polymeric samples were prepared with low density polyethylene LDPE (Carmel Chemicals Ltd.) by hot pressing with atomically flat heated plates. LDPE films with a thickness of 1mm manufactured by extrusion were sandwiched between aluminum foil and a steel computer hard disk with a diameter of 10 cm, which was atomically flat. Then the sandwich was pressed under pressure of 1.3 MPa and a temperature of $170^0$ C for 30 minutes. The sample was cooled at ambient conditions. The roughness of the pressed LDPE films was established with AFM as 5-10nm. The thickness of the LDPE films was 0.7±0.1 mm.

The LDPE films were exposed to a radiofrequency (13MHz) inductive air plasma discharge under the following parameters: pressure $P = 0.5$, 1.0 and 2.0 Torr, power 18 W under an ambient temperature. The details of the experimental setup are supplied in Ref. 16. For every pressure value, series of measurement of the contact angle, which is supposed to be connected to the surface charge density, was carried out for several exposition times: 0.05, 0.10, 0.15, 0.30, 0.50, 0.70, 1.00, 3.00 and 10.0 s.

Contact angles were measured with the Ramé-Hart Advanced Goniometer Model 500-F1. A series of 10 experiments was carried out for every sample. The results were averaged.

## 3. Results and discussion

### 3.1. Experimental study of the interaction of cold plasma with polymer surfaces

It is well-known that plasma treatment strongly modifies (increases) the surface energy of polymer materials [1-4]. The natural macroscopic measure of the surface energy is the apparent contact angle [30-31]. The accurate experimental establishment of the contact angle turns out to be a challenging task, due to the phenomenon of the contact angle hysteresis [32-37]. However, for atomically smooth polymer surfaces, the effect of the contact angle hysteresis is not essential and may be neglected. Hence, for the characterization of plasma irradiated LDPE surfaces we used the so called "as placed" apparent contact angle (APCA), introduced recently by Tadmor *et al.* [37]. Thus, the measurement of the "as placed" APCA on plasma treated LDPE films (according the procedure described in the Experimental Section) supplies at least qualitative information about the processes occurring on the polymer surfaces, exposed to plasma. The kinetics of the change of APCA with the time of exposure is depicted in Fig. 2.

Time dependence of the contact angle on time $t$ was fitted by the following empiric expression:

$$\theta(t) = \tilde{\theta} \exp\left(-\frac{t}{\tau_{\exp}}\right) + \theta_{sat}, \qquad (1)$$

where $\theta$ is the APCA, $\theta_{sat}$ is the empiric saturation contact angle corresponding to the infinite time of plasma irradiation, $\tilde{\theta} + \theta_{sat}$ is the initial APCA, and $\tau_{\exp}$ is the experimentally established characteristic time calculated by fitting the experimental data with Exp. 1 (see Table 1). These characteristic times give a semi quantitative description of the effect of plasma treatment on LDPE films, indicating the time scales when the change in wettability comes to the saturation. Further changes of wettability will be related to the etching of the polymer surface, resulting in the change of its roughness [18]. However, the effect of etching is beyond the scope of our manuscript.

### 3.2. Physical model describing the interaction of cold plasma with polymer surfaces

Consider the experimental situation depicted in Fig. 3, namely ions accelerated by the electric field of the sheath colliding with solid or liquid polymer surfaces. It is well-known that the condensed organic matter exposed to the cold plasma is markedly modified (hydrophilized) [1-5, 23]. A broad diversity of physical events occurs under "bombardment" of a polymer surface by cold plasma ions, including elastic collisions, inelastic collisions, recombination of ions and orientation of dipole groups constituting a polymer substrate with the electric field of the plasma sheath [29, 38].

Consider the interaction of dipole groups of polymer chains with the electrical field of the sheath. Assume that the surface is built of moieties possessing the dipole moment $\vec{p}$. It seems reasonable to partially relate the hydrophilization of the surface to the orientation of these moieties by the electric field of the sheath $\vec{E}$ (see Fig. 3). The dimensionless parameter $\varsigma$, describing the interaction of the dipoles constituting the surface with the electric field of the sheath $\vec{E}_{sh}$, is given by [39]:

$$\varsigma = \frac{pE_{sh}}{k_B T}. \qquad (2)$$

The upper value of the achievable electric field of the sheath may be estimated as $E_{sh}^{max} \cong \frac{\Phi_W}{\lambda_{De}} \cong \frac{100\,\text{V}}{10^{-4}\,\text{m}} = 10^6 \frac{\text{V}}{\text{m}}$, where $\Phi_W \approx 100V$ is the potential of the wall, and $\lambda_{De} \approx 10^{-4}\,\text{m}$ is the Debye length of the cold plasma [29]. Substituting $p \cong 1\text{D} \cong 3.3 \times 10^{-30} \text{C} \cdot \text{m}$, which is typical for moieties constituting polymer surfaces [40], we obtain the upper estimation of $\varsigma$ for room temperatures: $\varsigma \cong 10^{-3}$. This means that the observed hydrophilization of organic surfaces by cold plasmas could hardly be related to orientation of the dipole moieties forming the surface by the electrical field of a plasma sheath. This orientation will rapidly be destroyed at ambient conditions by the thermal agitation of dipole groups. Thus, the pronounced hydrophilization of polymer surfaces by cold plasma could hardly be related to the orientation of polar groups by the electrical field of the sheath. It should be stressed that the orientation of dipole polymer moieties will be essential when these moieties are exposed to intermolecular forces, as shown in Ref. 41.

Elastic collisions and recombination also hardly contribute to the hydrophilization of the polymer surface by cold plasma. Consider inelastic collisions of ions with the substrate, resulting in the charging of a polymer surface [42-43]. These inelastic collisions may proceed according to the mechanism developed in the classical works of Su and Bowers [44-45] who treated in detail the collisions of ions with polar molecules and quantified the capturing of ions by polar molecules. We assume that the pronounced hydrophilization of organic surfaces is at least partially related to capture of ions by polar groups constituting the solid (or perhaps liquid) surfaces.

From the experimental data reported in the previous sections and supported by the results of other groups [46-48], we know that APCAs come to saturation under plasma treatment within the time domain of 1-5 s. It is plausible to suggest that within this interval the electrical charging of the surface stops, and the surface charge density gains its saturation value. The charging stops when the modulus of the electrical field produced by the charged polymer surface $E_{surf}$ attains the value of the electrical field of the plasma sheath, i.e. the Condition (3) takes place:

$$E_{surf} = E_{sh} \quad . \tag{2}$$

When this equilibrium condition occurs, the net force accelerating ions towards the polymer surface becomes zero. The electric field produced by the polymer surface charged by ions is given by:

$$E_{surf} = \frac{\sigma(t)}{2\varepsilon_0}, \qquad (3)$$

where $\sigma(t)$ is the time-dependent surface charge density of the polymer charged by plasma. Thus, the maximal attainable surface charge density $\tilde{\sigma}$ is given by:

$$\tilde{\sigma} = 2\varepsilon_0 E_{sh} \qquad (4)$$

The kinetics of charging a polymer surface by plasma may be described by the following phenomenological equation:

$$\frac{dN(t)}{dt} = (N_0 - N(t))\alpha v n S, \qquad (5)$$

where $N(t)$ is the surface density of the sites which trapped ions (CH$_2$ groups in the case of LDPE studied in the paper), $v$ is the velocity of the ions, $N_0$ is the initial surface density of all surface sites which may trap ions, $n$ is the volume density of plasma ions (which depends on the plasma pressure $n\sim P$ [29]), $S \cong 2 \cdot 10^{-19} m^2$ is the surface area per one CH$_2$ group (see Ref. 40), and $\alpha$ is the phenomenological coefficient, describing the percentage of trapped ions, giving rise to the charging of the surface. The relation between $\sigma(t)$ and $N(t)$ is obvious:

$$\sigma(t) = eN(t), \qquad (6)$$

where $e$ is the electron charge, and it is supposed, that one site traps only one ion. Introducing $\sigma_0 = eN_0 = \frac{e}{S} \cong 0.8 C \cdot m^{-2}$, which is the maximal surface charge density corresponding to the total filling of trapping sites (unattainable in our experimental situation), we have:

$$\frac{d\sigma(t)}{dt} = (\sigma_0 - \sigma(t))\alpha v n S. \qquad (7)$$

The integration of Eq. 7 yields:

$$\sigma(t) = \sigma_0(1 - \exp[-\alpha v n S t]). \qquad (8)$$

Assuming that the maximal attainable surface density of the electric charge $\tilde{\sigma}$ is gained in time $t = \tilde{\tau}$, we obtain:

$$\tilde{\sigma} = \sigma(t = \tilde{\tau}) = \sigma_0(1 - \exp[-\alpha v n S \tilde{\tau}]). \qquad (9)$$

Combining Exps. 4 and 9 enables the calculation of the phenomenological coefficient $\alpha$ according to:

$$\alpha = -\frac{\ln(1-\frac{\tilde{\sigma}}{\sigma_0})}{vnS\tilde{\tau}} = -\frac{\ln(1-\frac{2\varepsilon_0 E_{sh}}{\sigma_0})}{v_B nS\tilde{\tau}} \quad . \tag{10}$$

For the sake of a rough estimation, we assume that: $\sigma_0 = 0.8\frac{C}{m^2}$; and $v_0 = v_B = 2 \div 6 \cdot 10^3 \text{m/s}$, which is the so-called Bohm velocity [29]. For the typical sheath field we infer $E_{sh} \cong 300\frac{V}{m}$ (which is much smaller than $E_{sh}^{\max}$ discussed above, see Ref. 29); $\tilde{\tau}$ and $n$ are on the order of magnitude of the values supplied in Table 1. The accurate measurement of the concentration of ions in the closed plasma chamber $n$ presents essential difficulties [49]. We established the concentration $n$ with a novel probe according to the protocol, described in detail in Ref. 50. The values of ion concentration corresponding to various values of the plasma pressure are also supplied in Table 1. Thus, the rough estimation of $\alpha$ yields: $\alpha \cong 10^{-8} \div 10^{-11}$, depending on the ion concentration. The value of $\tilde{\sigma}$ may be estimated according to the Exp. 9 as: $\tilde{\sigma} \cong 5.0\frac{nC}{m^2}$, and this surface charge density is much smaller than the maximal experimentally achievable charge density reported for polymer films in the literature [51].

It is convenient to rewrite the solution of the kinetic equation given by Exp. 8 in the following form:

$$\sigma(t) = \sigma_0(1-(1-\frac{\tilde{\sigma}}{\sigma_0})^{t/\tilde{\tau}}) \quad . \tag{11}$$

Since $\sigma_0 \gg \tilde{\sigma}$,

$$\sigma(t) = \tilde{\sigma}\frac{t}{\tilde{\tau}} \quad . \tag{12}$$

Exps. 11-12 enable prediction of the time evolution of the surface density of the electrical charge. They comprise physical quantities, which may be calculated (such as $\sigma_0$ and $\tilde{\sigma}$) or established experimentally (such as $\tilde{\tau}$), as shown in the previous Section. Obviously they are applicable when $t < \tilde{\tau}$ takes place.

The characteristic times $\tilde{\tau}$, at which the polymer surface gains its maximal charge, obviously do not coincide with the characteristic times $\tau_{exp}$, obtained from the measurements of APCAs and supplied in Table 1; but it plausible to suggest that they are closely correlated. We did not enter into details of the microscopic events occurring under cold plasma/polymer surface interaction; however, it is reasonable to suggest that these times will be reciprocal to the concentration of ions $n$. This plausible suggestion is supported by the data supplied in Table 1.

**Conclusions**

The interaction of cold radiofrequency plasma with low-density polyethylene substrates was studied experimentally and theoretically. Study of the wettability of the plasma treated polyethylene substrates yielded the estimation of the characteristic times at which the effect of plasma treatment comes to saturation. We related the effect of saturation of the contact angle to the saturation resulting from charging of a polymer surface by the plasma.

The interaction of cold radiofrequency plasma with condensed surfaces includes a series of complicated physico-chemical events. We supposed that one of these events is the trapping of ions by $CH_2$ groups constituting the surface of polyethylene, accompanied by the charging of the polymer surface. We conjectured that the charging stops when the modulus of the electrical field produced by the charged polymer surface attains the value of the electrical field of the plasma sheath. A phenomenological kinetic model of the electrical charge trapping is presented. The solution of the kinetic equation enabled an estimation of the phenomenological coefficient of ion trapping by the surface. The characteristic time at which the effect of plasma treatment comes to the saturation is inversely proportional to the concentration of ions $n$. We conclude that the trapping of ions by polar groups of polymer surface, resulting in its electrical charging, presents one of the important mechanisms of interaction of cold plasmas with organic surfaces.

**Acknowledgements**

We are thankful to Mrs. Y. Bormashenko for her help in preparing this manuscript. We are thankful to Mr. A. Shaposhnikov for numerous useful discussions.

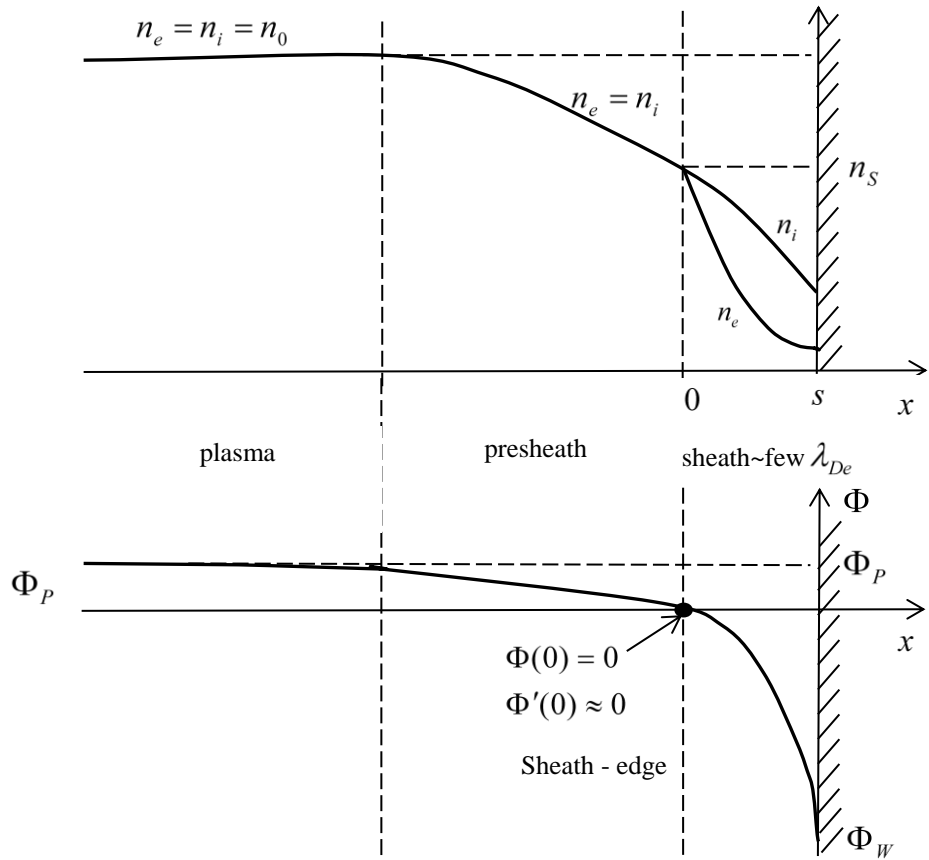

Fig. 1. Scheme of a sheath in a contact with the wall for the DC plasma; $n_e; n_i$ are the concentrations of electrons and ions respectively, $\Phi$ is the potential; $\Phi_P; \Phi_W$ are the potentials of the plasma and wall respectively; $\lambda_{De}$ is the Debye length.

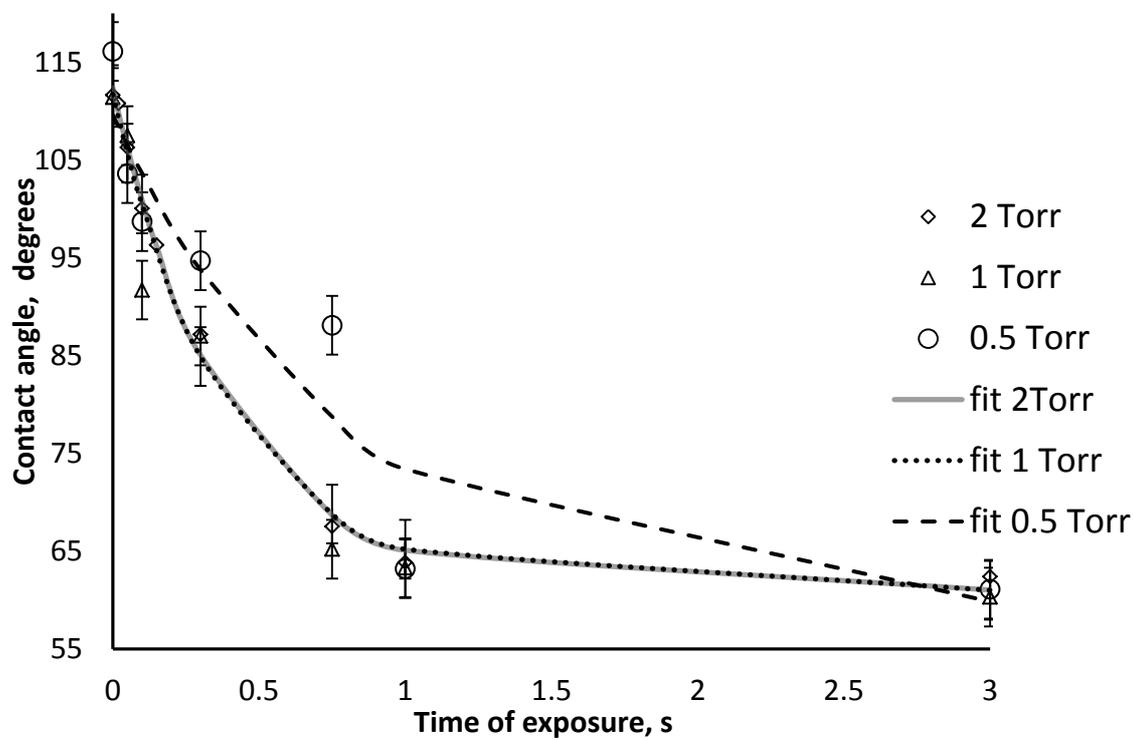

Fig. 2. The kinetics of the change of apparent contact angle with the time of exposure of LDPE film to plasma.

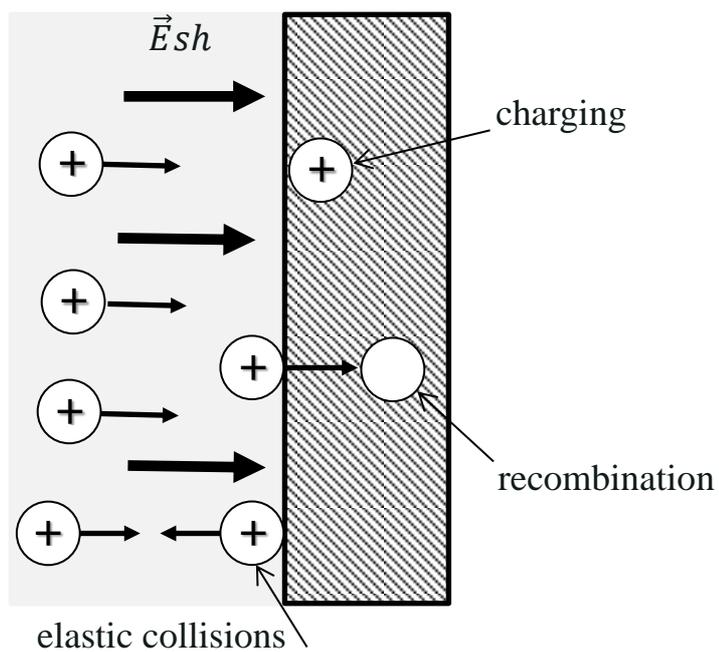

Fig. 3. Scheme depicting various events occurring under collisions of plasma ions accelerated by the sheath field $\vec{E}_{sh}$ with the polymer surface: elastic collisions, recombination and trapping of ions, accompanied by the electrical charging of the surface.

Table 1. Parameters of plasma (pressure and concentration of ions), and corresponding calculated and experimentally established characteristic times of interaction of cold plasma with a polymer substrate.

| $P$ (Torr) | $n$, molecule/m$^3 \times 10^{-15}$ | $\tau_{exp}$, s |
|---|---|---|
| 0.5 | 0.025÷0.25 | 0.9 |
| 1.0 | 0.1÷1.0 | 0.5 |
| 2.0 | 20÷200 | 0.34 |